\documentclass[preprint]{emulateapj}
\usepackage{apjfonts}

\slugcomment{ApJ Letters accepted}

\shorttitle{Identification of BHB stars}

\begin{document}

\title{The Identification of BHB stars using \emph{GALEX} and other photometry} 

\author{T.D.Kinman\altaffilmark{1,*},
S. Salim \altaffilmark{1,*} and 
L. Clewley\altaffilmark{2}}

\altaffiltext{1}{NOAO, P.O. Box 26732, Tucson, AZ 85726 \\
$^{\star}$ The NOAO is operated by AURA Inc., under cooperative 
agreement with the National Science Foundation.}
\altaffiltext{2}{Astrophysics, Department of Physics, Denys Wilkinson Building,
Keble Road, Oxford OX13RH}

\begin{abstract}

Halo samples that have no kinematic bias have been successfully isolated by 
photometric observations.
We show that the $NUV$ magnitude of the \emph{GALEX} All-Sky Survey can be used, 
together with Johnson $BV$ magnitudes in a 
($NUV - V)_{0}$ {\it vs.} $(B-V)_{0}$ plot, to distinguish
blue horizontal branch (BHB) from other 
 A stars of the same $(B-V)$ color for 12$<$$V$$<$18. In addition, we use
 SDSS $gr$ magnitudes  in a 
 $(NUV - r)_0$ {\it vs.} $(g-r)_0$ plot for 14$<$$r$$<$18. 
 The faint limit will be extended by $\sim$3 magnitudes for the 
  \emph{GALEX} Medium-Deep Survey. 
Attempts to use $NUV$ in conjunction with 2MASS magnitudes and ROTSE $m_r$ 
magnitudes did not prove useful. The 
($NUV - V)_{0}$ {\it vs.} $(B-V)_{0}$ plot was used to examine BHB star 
candidates of varying quality near the South Galactic Pole. We conclude 
that the addition of the \emph{GALEX} $NUV$ significantly adds to the 
reliabilty with which these stars can be identified.

\end{abstract}


\section{Introduction}

Blue horizontal branch (BHB) stars together with RR Lyrae stars are
 the most-used probes of the Galactic halo.  BHB stars can be
 distinguished from other blue stars (e.g. blue stragglers) by the
 width and shape of their Balmer lines and by the size of their Balmer
 jump. They have been discovered by using objective-prism spectroscopy
 (e.g.  Sanduleak  1988; Beers et al.\  1988;  Christlieb et
 al.\ 2005) or photoelectric photometry (e.g. Philip 1967; Pier
  1982,1983,1984; Sommer-Larsen et al.\ 1989; Preston et al.\  1991;
 Kinman et al.\ 1994; Clewley et al.\ 2004; Sirko et al.\ 2004; 
 Beers et al.\  2007) either alone or in combination with slit
 spectroscopy.  Preston et al.\ (1991) found that a separation of BHB
 from other A-stars in a $(U-B)_{0}$ vs. $(B-V)_{0}$ plot is only reliable in
 the color range $-$0.03$\leq$$(B-V)_0$$\leq$0.18 and, because of
 observational error, the classification is often ambiguous for stars
 near the ends of this range. Only in this color range can the Balmer jump 
 and the shape and widths of the Balmer lines be used to distinguish the
 BHB stars (e.g. Clewley et al.\ 2002).
 Improved classification can also come from using high S/N high
 resolution spectra but this has only been achieved so far for the
 brighter nearby BHB (Kinman et al.\ 2000; Behr 2003).  One can
 also improve the photometric selection by using more than one
 photometric system. We here discuss the appropriate colors for
 identifying BHB stars in a variety of photometric systems with
 particular reference to the new UV photometry from the Galactic
 Explorer Satellite (\emph{GALEX}, Martin et al.\ 2005).

\section{The color windows for BHB stars using  \emph{GALEX} magnitudes.}

\subsection{\emph{GALEX} and Johnson BV photometry.}

The \emph{GALEX} surveys give a far-UV
magnitude ($FUV$) and a near-UV magnitude ($NUV$) at effective wavelengths
1516 and 2267 \AA \ respectively; these are on the $AB$ system.   
These surveys become non-linear for stars brighter than 14 to 15th
magnitude (Morrissey et al.\ 2007), and so require faint BHB stars for
calibration. A suitable BHB sample (11.4$<$$V$$<$16.5) near the
NGP (where the extinction is low) is discussed in Kinman et
al.\ (2007).  The coordinates, photometry, radial velocities and
abundances of these stars are to be given in T.D.\ Kinman et al.\ (2007, 
in prep.)
which will also list the blue non-BHB stars in the same fields. The
techniques for selecting these stars are described in Kinman et
al.\ (1994, 1996). 
 Of this NGP sample, 31 BHB stars are in the
\emph{GALEX} releases GR2 and GR3, 40 in the SDSS (DR5) survey (York et
al.\ 2000; Adelman-McCarthy et al.\  2007) and 15 are in both surveys.
 Thirty nine non-BHB stars in the same NGP field were also
identified in the \emph{GALEX} releases.  
One BHB star and five non-BHB stars with $NUV$ $\leq$ 14.0 (corresponding
roughly to $V$ = 12.0 for $(B-V)$ $\sim$ 0.0) were excluded in order
to avoid non-linearity effects.  $FUV$ and $NUV$ were corrected for
interstellar extinction using A$(FUV)$ = 8.16 E$(B-V)$ and A$(NUV)$ =
8.90 E$(B-V)$ (Rey et al.\ 2006); the E$(B-V)$ were taken from Schlegel
et al.\ (1998).

\begin{figure}
\epsscale{1.0}
\plotone{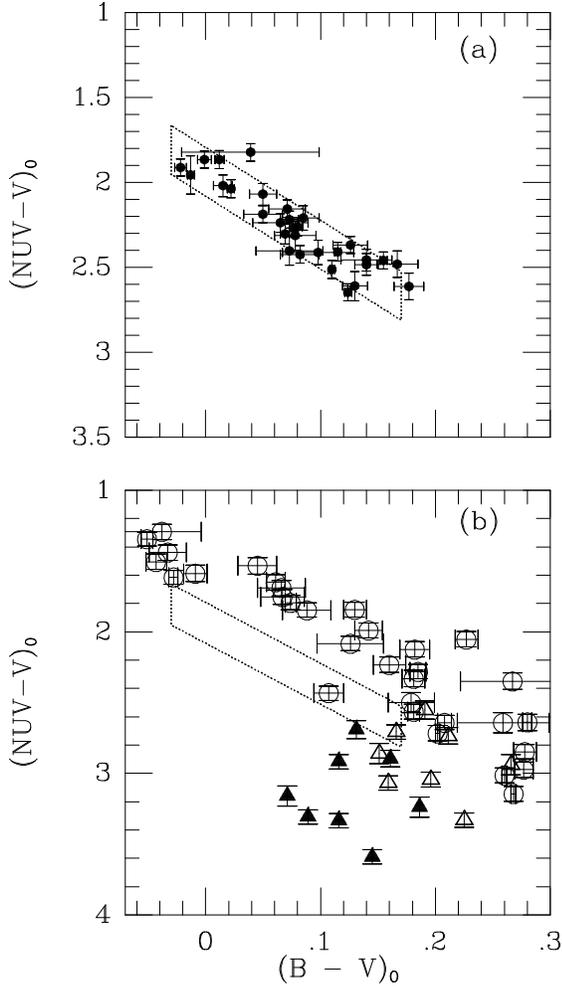}
\caption{$(NUV - V)_{0}$ {\it vs.} $(B-V)_{0}$ for (a) BHB stars (filled circles)
 and (b) non-BHB stars (open circles) and c-type RR Lyrae stars (open triangles
at mean light, filled triangles at maximum light). The $(B-V)_{0}$, $(NUV - V)_{0}$
coordinates of the corners of the parallelogram that encloses the BHB stars are:
$-$0.03, 1.665; $-$0.03, 1.953; +0.17, 2.814; +0.17, 2.526.  
\label{Fig1}}
\end{figure}

The photometric separation of BHB from other A stars can be achieved
by plotting a gravity-sensitive indicator, such as $(U-B)$, against a
temperature indicator, such as $(B-V)$. We find that 
 ($NUV - V)_{0}$ and ($NUV -B)_{0}$ are effective surrogates for the 
gravity-sensitive $(U-B)$. All photometric 
classifications are limited to an  effective magnitude range.
At the faint end the photometric errors are too large for stars of
different types to be separated; at the bright end the detector becomes
non-linear.
We do not use the $FUV$ here because it has a small effective range 
for stars in the All-Sky Survey (AIS); it may prove useful when the deeper 
\emph{GALEX} surveys are more complete.
 ($NUV - V)_{0}$ and ($NUV -B)_{0}$ 
are similar as gravity discriminators but the former includes
fewer c-type RR Lyrae stars and was therefore adopted.  A plot of
($NUV - V)_{0}$ {\it vs.} $(B-V)_{0}$ (Fig. 1(a)) shows that the BHB
stars (filled circles) lie in a compact area (the dotted
parallelogram) whereas the non-BHB stars (open circles) largely lie
outside this area (Fig. 1(b)). Eight c-type RR Lyrae stars are also shown. 
 We do not know the epochs of the \emph{GALEX} data  and so 
 the phases of these variables at the times of their \emph{GALEX}
observations are not known.  We therefore plotted them both for the
case where the variable is at maximum light (filled triangles) and
where it is at mean light (open triangles). Confusion of these
variables with BHB stars is possible on this plot and so 
the reddest BHB star candidates should be checked for variability.  Most of
our \emph{GALEX} data comes from the AIS for which the error
in $NUV$$\sim$0.15 mag at $NUV$ = 20.0;  the separation of BHB from
non-BHB stars will become increasingly inefficient for stars fainter
than this. The effective useful range of the AIS data is therefore
14.0$<$$NUV$$<$20.0 which corresponds roughly to 12.0$<$$V$$<$18.0 for
BHB stars.  A likely practical use of the $(NUV - V)_{0}$ {\it vs.}
$(B-V)_{0}$ plot is to provide improved selection where Johnson
photometry is available (e.g.  Beers et al.\ 2007). An example of this
is given in Sec 2.4.

\subsection{\emph{GALEX} and SDSS photometry.}

\begin{figure}
\epsscale{1.0}
\plotone{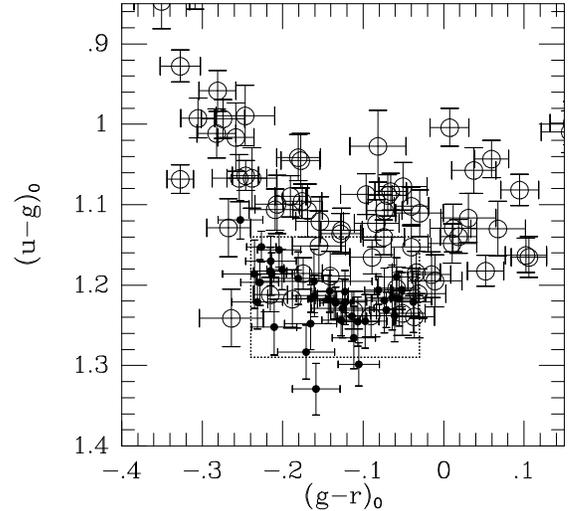}
\caption{  $(u-g)_{0}$ {\it vs.} $(g-r)_{0}$ (SDSS colors)  
      for BHB stars (filled circles) 
      and non-BHB stars (open circles). The dotted rectangle encloses  the 
      area occupied by the BHB stars. 
\label{Fig2}}
\end{figure}

\begin{figure}
\epsscale{1.0}
\plotone{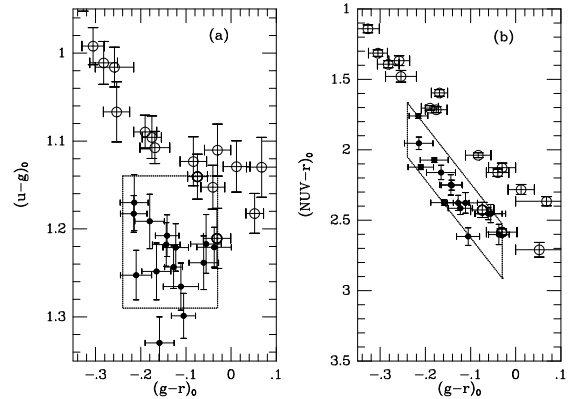}
\caption{ (a) $(u-g)_{0}$ {\it vs.} $(g-r)_{0}$ (SDSS colors)  and (b)
      $(NUV- r)$~{\it vs.} $(g-r)_{0}$ for stars with GALEX magnitudes (same
      symbols as Fig. 2). The classification of the two stars with bold 
      ircles is ambiguous from the photometry.
      The dotted parallelograms enclose the 
      area occupied by the BHB stars. The $(g-r)_{0}$, $(NUV - r)_{0}$ 
      coordinates of the corners of the parallelogram in (b) are $-$0.24, 1.665; 
      $-$0.24, 2.050; $-$0.03, 2.911; $-$0.03, 2.526.
\label{Fig3}}
\end{figure}

Sirko et al.\ (2004) have selected BHB stars in the SDSS system by
 using a $(u-g)_0$ {\it vs.} $(g-r)_0$ plot together with
 spectroscopic criteria.  The BHB stars in our NGP sample in the same
 color plot (Fig 2) lie in the ranges 1.14$\leq$$(u-g)_0$$\leq$1.28
 and $-$0.24$\leq$$(g-r)_0$$\leq$$-$0.04 (as shown by the box).  The
 color-window ($-$0.31$\leq$$(g-r)_0$$\leq$$-$0.13) given in Fig. 8 of
 Sirko et al.\  (2004) is significantly bluer than this and corresponds
 roughly to $-$0.09$\leq$$(B-V)_{0}$$\leq$$+$0.10. In both cases,
 however, the region occupied by the BHB stars (filled circles) is heavily
 contaminated by non-BHB stars (open circles). Fig 3(a) shows the 
 $(u-g)_0$ {\it vs.} $(g-r)_0$ plot and Fig 3(b) shows the
 $(NUV - r)_0$ {\it vs.} $(g-r)_0$ plot (with the same symbols) for the 
 subset of stars in Fig. 2 that have reliable \emph{GALEX} data. 
 The lower contamination of the BHB star region by non-BHB stars in Fig 3(a)
 compared to Fig 2, may be a fluctuation produced by the smaller number of
 stars in Fig 3(a) and/or some selection effect produced by the 
 \emph{GALEX} survey itself. The presence of two non-BHB stars (shown as 
 bold open circles) in the BHB star region in both Fig 3(a) and Fig 3(b)
 suggests an ambiguity in their classification. Otherwise, the use of
 $(NUV - r)_0$ clearly gives an effective and \emph{independent} method 
 of separating the BHB from non-BHB stars. The use of both SDSS and 
 \emph{GALEX}
 surveys materially adds to the accuracy of the classification process.
 Classification  should be possible from the
 \emph{GALEX} AIS  survey for stars with 14.2$<$r$<$18.0; the bright
 limit is set by saturation in the SDSS data and the faint limit by
 the errors in NUV.  The use of the \emph{GALEX} Medium-Deep (MIS) survey, 
 which largely overlaps with the SDSS, 
 should extend the faint limit by $\sim$3 mag.

\subsection{\emph{GALEX} and other photometry.}

\begin{figure}
\epsscale{1.0} 
\plotone{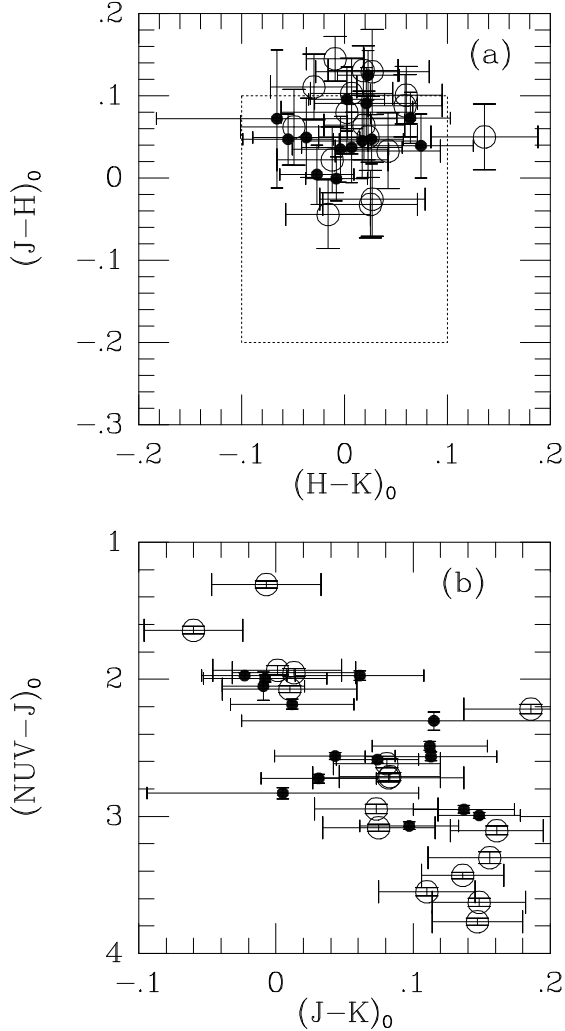}
\caption{ 
          (a) 2MASS $(J-H)_{0}~{\it vs.}~(H-K)_{0}$  for BHB stars (filled 
      circles) and non-BHB stars (open circles) at the NGP. The dotted rectangle
     shows the selection window used by Brown et al.\ (2004).
          (b)  $(NUV - J)_{0}~{\it vs.}~(J-K)_{0}$  for BHB stars (filled 
      circles) and non-BHB stars (open circles) at the NGP. 
\label{Fig4}}
\end{figure}

\begin{figure}
\epsscale{1.0}
\plotone{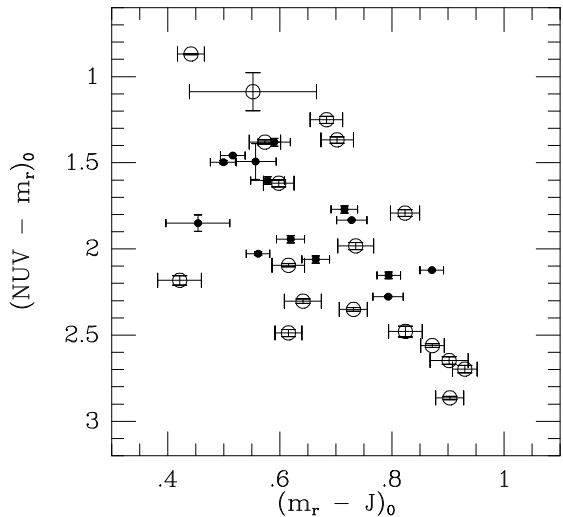}
\caption{ 
   $(NUV - m_r)_{0}$~{\it vs.} $(m_r - J)_{0}$ where $NUV$ is from {\it GALEX}, $J$ from 
   2MASS and $m_{r}$ is from the NSVS. The filled and open circles show BHB
   and non-BHB stars respectively from the NGP field.
\label{Fig5}}
\end{figure}

It is desirable to find an all-sky survey whose data could be used in
 conjunction with \emph{GALEX} to identify the brighter BHB stars ($V < 14$).
 Brown et al.\ (2004) suggested that 2MASS data can be used to identify BHB stars in a 
 $(J-H)_{0}$ {\it vs.} $(H-K)_{0}$ plot. Fig 3(a) shows our NGP sample of
 BHB and non-BHB stars in this plot. It is seen that their
 color-window (shown by the dotted rectangle) is not optimal and that
 there is little separation between the two types of star.  It is
 found (Fig 3(b)) that the $(NUV - J)_{0}$ {\it vs. }  $(J-K)_{0}$
 plot is also ineffective since $(NUV - J)_{0}$ is not a possible
 gravity indicator. This is also true of $(NUV - m_{r})$ (where
 $m_{r}$ is the unfiltered CCD magnitude of the Northern Sky
 Variability Survey (NSVS) (Wozniak et al.\ 2004))\footnote{ This
 survey covers the sky North of declination $-$38$\arcdeg$.}.  Fig 4
 shows the plot of $(NUV - m_{r})$ {\it vs.} $(m_r - J)$ for our NGP
 stars; only a weak separation between the BHB and non-BHB stars is
 seen (probably because $m_{r}$ has such a broad bandwidth).  The $V$ magnitudes
 of the All Sky Automatic Survey (ASAS) (Pojmanski 2002) are available for 
 declinations South of +28$\arcdeg$, but the magnitude range in which both
 these $V$ and $NUV$ are accurate is rather limited. Apart from this, therefore,
 there does not seem to be a way of identifying the brighter BHB
 stars in currently published wide-field surveys.

\subsection{\emph{GALEX} and a sample at the South Galactic Pole (SGP).}

\begin{figure}
\epsscale{1.0}
\plotone{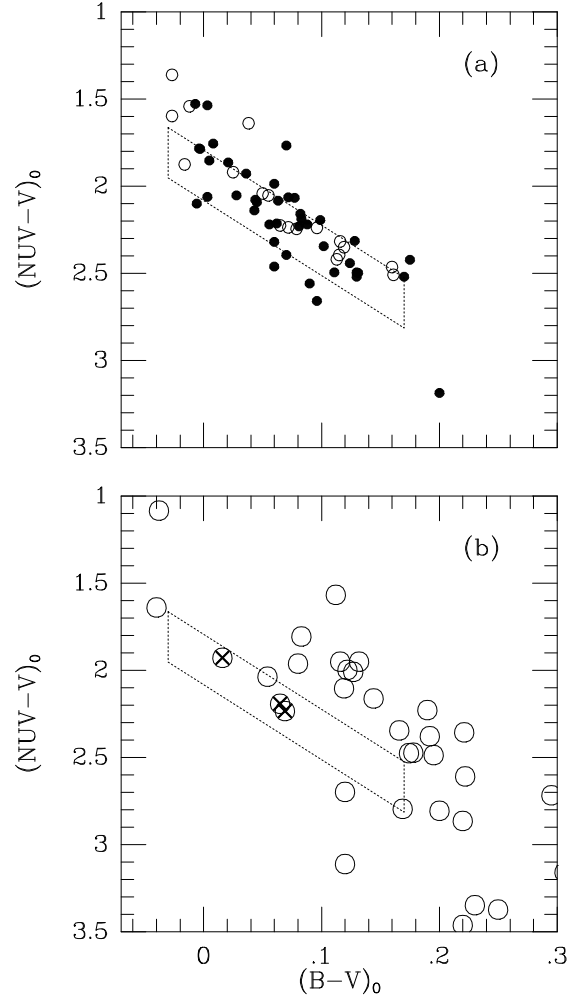}
\caption{$(NUV - V)_{0}$~{\it vs.} $(B-V)_{0}$ for (a) Most likely (filled 
  circles) and quite likely (small open circles) BHB star candidates near 
  the SGP. (b) Least likely candidates shown as large open circles. The dotted
  rectangle is that which encloses the BHB stars in a similar plot for the
  NGP BHB stars. The crosses are explained in the text.
\label{Fig6}}
\end{figure}

An (unpublished) catalog of BHB candidates within 10$\arcdeg$ of the
SGP has been compiled from various sources (e.g. Christlieb et
al.\ 2005; Beers et al.\ 1988; Lance, 1988). The data is of varying
accuracy and (largely from their UBV colors) they were judged to be
(a) most likely (b) quite likely (c) not likely to be BHB stars. The
first two classes are shown by filled and small open circles
respectively in the plot of ($NUV - V)_{0}$ {\it vs.} $(B-V)_{0}$
(Fig. 5(a)) where the dotted rectangle is that which encloses the NGP
BHB stars in Fig 1(a).  The difference between the distributions of the
filled and open circles is not marked and with the exception of a few
outliers, most of these stars probably are BHB stars.  We suspect that
the scatter is largely caused by the (unknown) errors in $(B-V)_{0}$
in this heterogeneous data. There may also be systematic differences
between the Northern and the Southern hemisphere photometry standards
that affect $(B-V)_{0}$ (Menzies et al.\ 1991). The classification of
the outliers in this plot therefore requires an assesment of the
likely errors in each $(B-V)_{0}$ and this is beyond the scope of this
Letter.  A similar plot for the SGP stars that were judged least
likely to be BHB stars is shown in Fig 5(b). A few lie inside the
rectangle that encloses the NGP BHB stars. The three marked with
crosses are BPS CS 22882-32 and BPS CS 22942-34 whose radial
velocities are +98 and +192 km s$^{-1}$ (Pier  1983) respectively and PS 2-58 
whose radial velocity is +200 km s$^{-1}$ (Rodgers  1971); these high radial 
velocities also support the classification of these stars as BHB stars. 
The efficiency in separating BHB stars from their interlopers using plots
such as  ($NUV - V)_{0}$ {\it vs.} $(B-V)_{0}$ 
 is dependent on the accuracy of the
color used as the temperature indicator. Consequently, it would be advantageous
to use a temperature indicator that combines the information from all 
the optical and infrared magnitudes that are available.

\section{Summary}

It is shown that the \emph{GALEX} $NUV$ magnitude can be used in conjunction 
with Johnson $BV$ magnitudes in a 
($NUV - V)_{0}$ {\it vs.} $(B-V)_{0}$ plot to distinguish BHB from other
 A stars of the same $(B-V)$ color. A similar separation can made using $NUV$
and the SDSS $gr$ magnitudes in a 
 $(NUV - r)_0$ {\it vs.} $(g-r)_0$ plot. These plots are effective for stars in
the \emph{GALEX} AIS  in the 
magnitude ranges 12$<$$V$$<$18  and 
14$<$$r$$<$18 for Johnson and SDSS colors respectively. The faint limit is 
 $\sim$3 magnitudes fainter for the \emph{GALEX} Medium-Deep Survey. The bright 
limit is set by the onset of non-linearity in the \emph{GALEX} detector in the 
case of the Johnson colors and by the saturation of the SDSS detectors in their 
case. Attempts to use $NUV$ in conjunction with 2MASS magnitudes and ROTSE $m_r$
magnitudes to  separate  BHB from non-BHB stars were not successful.
The ($NUV - V)_{0}$ {\it vs.} $(B-V)_{0}$ plot was used to confirm BHB star 
candidates of varying quality near the South Galactic Pole. It is concluded 
that the addition of the \emph{GALEX} $NUV$ to existing data gives a simple and 
powerful way of identifying these stars.

\acknowledgments We thank Heather Morrison for help in selecting SDSS
   stars. We acknowledge NASA's support for the {\it GALEX}
   mission. Funding for SDSS has been provided by the Alfred
   P. Sloan Foundation, the Participating Institutions, NSF, DoE,
   NASA, Monbukagakusho, and the Max Planck Society. This publication
   makes use of 2MASS, which is a joint project
   of the University of Massachusetts and the IPAC/Caltech, funded by
   NASA and NSF. Finally, we thank the referee for suggestions which
   materially clarified parts of the paper.


\begin{thebibliography}{}

\bibitem[Adelman-McCarthy et al. 2006]{ade06}
Adelman-McCarthy, J.~K., et al.\ 2006, \apjs, 162, 38
\bibitem[Beers, et al., 1988]{bee88} Beers, T.C., Preston, G.W., Shectman, S.A.,
     1988, ApJS, 67, 461  
\bibitem[Beers et al., 2007]{bee07}Beers, T.C., Almeida, T., Rossi, S. et al.,
 2007, ApJS, 168, 277  
\bibitem[Behr, 2003]{beh03}Behr, B.B.  2003, ApJS, 149, 101  
\bibitem[Brown et al. 2004]{bro04}Brown, W.R., Geller, M.J., Kenyon, S.J.,
      et al.  2004, AJ, 127, 1555 
\bibitem[Christlieb et al. 2005]{chr05}Christlieb, N., Beers, T., Thom, C.,
       et al. 2005, A\&A, 431, 143  
\bibitem[Clewley, et al. 2002]{cle02}Clewley, L., Warren, S.J., Hewett, P.C., et al.  
     2002, MNRAS, 337, 87 
\bibitem[Clewley, et al. 2004]{cle04}Clewley, L., Warren, S.J., Hewett, P.C., et al.  
     2004, MNRAS, 352, 285  
\bibitem[Kinman et al.,  1994]{kin94} Kinman, T.D., Suntzeff, N.B., Kraft, R.P.,
 1994, \aj, 108, 1722 
\bibitem[Kinman et al.,  1996]{kin96} Kinman, T.D., Pier, J.R., Suntzeff, N.B. 
 1996, \aj, 111, 1164 
\bibitem[Kinman et al., 2000]{kin00} Kinman, T.D., Castelli, F., Cacciari, C.,
   et al. 2000, A\&A, 364, 102  
\bibitem[Kinman, et al. 2007a]{kin07a} Kinman, T.D., Cacciari, C., Bragaglia A.,
   et al.  2007, MNRAS, 375, 1381 
\bibitem[Lance, 1988]{lan88} Lance, C. 1988, ApJS, 68, 463  
\bibitem[Martin et al., 2005]{mar05} Martin, D.C., Fanson, J., Schiminovich,
    D., et al.  2005, ApJ, 619, L1
\bibitem[Menzies et al. 1991]{men91} Menzies, J., Mareng, F., Laing J. et al. 
    1991, MNRAS, 248, 642
\bibitem[Morrissey et al. 2007]{mor07} Morrissey, P. et al. 2007,  
\apjs\ Submitted
\bibitem[Pier, 1982]{Pie82}Pier, J.R. 1982, AJ, 87,  1515  
\bibitem[Pier, 1983]{Pie83}Pier, J.R. 1983, ApJS, 53, 791  
\bibitem[Pier, 1984]{Pie84}Pier, J.R. 1984, ApJ, 281,  260   
\bibitem[Philip, 1967]{phi67}Philip, A.G.D. 1967, ApJ, 148, L143 
\bibitem[Pojmanski, 2002]{poj02} Pojmanski, G. 2002, Acta Astonomica, 52, 397
\bibitem[Preston et al. 1991]{pre91}Preston, G.W., Shectman, S.A., Beers, T.C.,  1991, ApJ, 375, 121  
\bibitem[Rey et al. 2006]{rey06} Rey, S.-C., Rich, R.M., Sohn, S.T., Yoon,
     S.-J., et al. 2006, astro-ph 0612203 
\bibitem[Rodgers, 1971]{rod71} Rodgers, A.W. 1971, ApJ, 165, 581  
\bibitem[Sanduleak. 1988]{san88}Sanduleak, N. 1988, ApJS, 66, 309  
\bibitem[Schlegel et al., 1998]{sch98} Schlegel, D., Finkbeiner, D. \& Davis,
 M., 1998, \apj, 500, 525 
\bibitem[Sirko et al., 2004]{sir04}Sirko, E., Goodman, J., Knapp, G.R., et al. 
      2004, AJ, 127, 899 
\bibitem[Sommer-Larsen et al. 1989]{som89}Sommer-Larsen, J. Christensen P.R.,
     Carter, D. 1989, MNRAS, 238, 225  
\bibitem[Wozniak et al. 2004]{woz04} Woznial,, P.R., Vestrand, W.T., 
          Akerlof, C.W., et al.  2004, AJ, 127, 2436  
\bibitem[York et al. 2000]{yor00}York, Adelman, J., Anderson, J.E., et al. 
   AJ, 120, 1579
\end{thebibliography}
\end{document}